\documentclass{aip-cp}

\usepackage[numbers]{natbib}
\usepackage{rotating,wasysym,amssymb}
\usepackage{graphicx}

\usepackage{color}

\begin{document}

\title{On the Role of Metastable States in Low Pressure Oxygen Discharges}

\author[aff1,aff2]{J. T. Gudmundsson\corref{cor1}}
\author[aff1]{H. Hannesd{\'o}ttir}

\affil[aff1]{Science Institute, University of Iceland,  Dunhaga 3, IS-107 Reykjavik, Iceland}
\affil[aff2]{Department of Space and Plasma Physics, School of Electrical Engineering, KTH -- Royal Institute of Technology, SE-100 44, Stockholm, Sweden}
\corresp[cor1]{Corresponding author: tumi@hi.is}

\maketitle

\begin{abstract}
We use the one-dimensional object-oriented particle-in-cell Monte Carlo collision code {\tt oopd1}
to explore the spatio-temporal evolution of the electron heating mechanism  in  a capacitively coupled oxygen discharge in the pressure range 10 -- 200 mTorr.  The electron heating is most significant 
in the sheath vicinity during the sheath expansion phase. 
We explore how including and excluding detachment by the singlet metastable states  
O$_2$(a$^1 \Delta_{\rm g}$) and O$_2$(b$^1\Sigma_{\rm g}^+$) influences  
the heating mechanism, the effective electron temperature and electronegativity, in the oxygen discharge.  
We  demonstrate that the detachment processes have a significant influence on the 
discharge properties, in particular for the higher pressures.  At 10 mTorr the time averaged  electron heating 
shows  mainly ohmic heating in the plasma bulk (the electronegative core) and at higher pressures there is no ohmic heating in the plasma bulk, that is electron heating in the sheath regions dominates.
\end{abstract}

\section{INTRODUCTION}

Capacitively coupled plasma (CCP) radio frequency discharges are frequently used 
for applications such as plasma etching or plasma enhanced chemical vapor deposition (PECVD) 
processes in integrated circuit manufacturing.
Impact by energetic electrons leads to dissociation and ionization of feedstock gas, and thus
 creation of reactive radicals and ions. 
These applications usually require  feedstock gases that are complex mixtures of reactive gases, often electronegative.  One such electronegative discharge  is the oxygen discharge.  
The oxygen discharge and its mixtures are of significance in various materials processing applications including
etching of polymer films,  ashing of photoresist, oxidation, and deposition of oxide films.
The oxygen chemistry is complicated due to the presence of metastable atomic and molecular species.
It is in particular the two low lying metastable molecular states designated by  $a^1\Delta_{\rm g}$ and $b^1\Sigma_{\rm g}^+$, which are located
0.98 and 1.627 eV above the ground state, respectively, that play an important role. It is well established that collisions with these metastable
states have in many cases larger cross sections and thus higher reaction rates than corresponding collisions with the
ground state molecule.

A volume averaged global model is an approach to get a detailed description
of the plasma chemistry in low pressure discharges. 
Global model studies show that the dominant species in the oxygen discharge is the oxygen molecule in the 
ground state, O$_2$(X$^3\Sigma_{\rm g}^-$), followed  by the oxygen atom in the ground state, O($^3$P).  The
singlet metastable states O$_2$(a$^1 \Delta_{\rm g}$) and O$_2$(b$^1\Sigma_{\rm g}^+$) and the
metastable atom O($^1$D) are also present in the plasma in significant amounts \citep{kiehlbauch03:660,gudmundsson04:2073,toneli15:325202}. 
Furthermore, a recent study shows that the O$_2$(b$^1\Sigma_{\rm g}^+$)
density can overcome the O$_2$(a$^1 \Delta_{\rm g}$) density in the pressure range
below 100 mTorr \citep{toneli15:325202}.  The O$_2^+$-ions are in majority among the positive ions and the O$^+$-ion
 density is much smaller and has a sharp decrease for pressures above 4 mTorr.  It is also seen that the negative ion 
density ratio  [O$^{\bf -}$]/[O$^{\bf -}_2$] is 5.3 at 1 mTorr, and  1.3 at 100 mTorr \citep{toneli15:325202}. 
Such global models studies neglect spatial variations in the plasma parameters as well as the 
kinetics of the discharge.   
They allow us to  explore the relative reaction rates for the creation and destruction of 
particular species over the pressure
range of interest and therefore which reactions are of significance.

To explore the kinetics of the discharge we apply particle-in-cell Monte Carlo collision (PIC/MCC) simulations. 
 In a PIC/MCC simulation the plasma is represented as a collection of macro-particles.
Equations of motion are solved for each macro-particle. The electric and magnetic fields are calculated self-consistently using
charge densities and currents produced by these macro-particles.
We use the {\tt oopd1} (object oriented plasma device for one dimension)
code to simulate the discharge. 
 It has one dimension in space and three velocity components for the particles (1d-3v). 
 The {\tt oopd1} code is supposed to replace the widely used {\tt xpdx1} 
series ({\tt xpdp1}, {\tt xpdc1} and {\tt xpds1}).
It was developed to  simulate various types of plasmas, 
including processing discharges, accelerators and beams.  
The code is written in  modular structure and it includes features such as relativistic kinematics
and different weights for different species.  
Earlier we developed a basic oxygen discharge model and benchmarked it against the well known
{\tt xpdp1} code \citep{gudmundsson13:035011}.  Later we explored adding the singlet metastable molecule
O$_2$(a$^1\Delta_{\rm g}$) and the metastable oxygen atom O($^1$D) to the reaction set \cite{gudmundsson15:035016}.
 There we demonstrated that the addition of the
singlet metastable molecule O$_2$(a$^1\Delta_{\rm g}$) has a significant influence on the discharge, 
such as on the electronegativity, the effective electron temperature, and the electron heating 
processes for a pressure of 50 mTorr. We found that it is, in particular, detachment by the metastable molecule 
O$_2$(a$^1\Delta_{\rm g}$) that has a
significant influence on the discharge properties. In a subsequent study we 
explored how the addition of the singlet metastable molecule
O$_2$(a$^1\Delta_{\rm g}$) modified the discharge properties in the pressure range 
10 -- 500 mTorr \citep{gudmundsson15:153302}.  
We found that at low pressure (10 mTorr), ohmic (collisional) heating in the
plasma bulk dominates and the heating mechanism evolves
through a phase where both ohmic heating in the bulk and
sheath oscillation (collisionless) heating  contribute, and at higher pressure
(50 -- 500 mTorr), the electron heating occurs almost solely in
the sheath region.  At the higher pressures, detachment by the
metastable singlet molecule O$_2$(a$^1\Delta_{\rm g}$)  has a significant influence 
on the electron heating process but at the lowest pressures detachment 
by O$_2$(a$^1\Delta_{\rm g}$) has only a small influence on the heating process. 
At low pressure, the electron energy probability function (EEPF) is convex
and as the pressure is increased the number of low energy
electrons increases and the number of higher energy electrons 
($>10$ eV) decreases, and the EEPF develops a concave
shape or becomes bi-Maxwellian \citep{gudmundsson15:153302}. This contradicts what is
observed for a capacitively coupled argon discharge where
the EEPF evolves from being concave at low pressure to
become convex at high pressure \citep{godyak90:996,vahedi93:273} and for the chlorine 
discharge where the EEPF is found to be convex in the pressure
range of 10 -- 100 mTorr \citep{huang13:055020}. In a more recent study we added the singlet metastable
oxygen molecule  O$_2$(b$^1\Sigma_{\rm g}^+$) to the reaction set \citep{hannesdottir16:055002}.
We find that including the singlet metastable O$_2$(b$^1\Sigma_{\rm g}^+$)
further decreases the ohmic heating in the electronegative core and the effective electron temperature in the bulk
region. The effective electron temperature in the electronegative core is found to be less than 1 eV
in the pressure range 50 -- 200 mTorr which agrees with recent experimental findings. 
Furthermore, we have added an energy-dependent secondary electron emission yield for the positive oxygen ions 
O$_2^+$ and O$^+$ as well as the neutrals O$_2$ and O  \citep{hannesdottir16:055002}.
We find that including an energy-dependent secondary electron emission yield for O$^+_2$-ions has a
significant influence on the discharge properties, including decreased sheath width and increased electron density  \citep{hannesdottir16:055002}.
Other PIC/MCC studies of the oxygen discharge have also indicated a significant role of the 
singlet metastable state  O$_2$(a$^1\Delta_{\rm g}$) and in particular the detachment by the 
  O$_2$(a$^1\Delta_{\rm g}$) on the overall discharge properties \citep{bronold07:6583,derzsi16:015004}.  
However, these studies did not  include a complete reaction set for creation and destruction of the metastable states and the metastables  are not treated kinetically.  

Here, we investigate the spatio-temporal electron heating
dynamics in a geometrically symmetric oxygen discharge  by 1d-3v
PIC/MCC simulations.  In particular
we explore the influence of the electron detachment from the negative ion O$^-$ by the  singlet molecular metastables 
 O$_2$(a$^1\Delta_{\rm g}$) and  O$_2$(b$^1\Sigma_{\rm g}^+$) on the  spatio-temporal electron heating
dynamics, the electronegativity, and the effective electron temperature.


\section{THE 
SIMULATION AND THE OXYGEN DISCHARGE}

We assume a  capacitively coupled discharge where one of the electrodes is driven by an rf voltage (left hand electrode)
while the other (right hand electrode) is grounded. 
We assume the discharge to be operated at a single frequency
 of 13.56 MHz, and voltage amplitude of  $V_0 = 222$ V with an electrode
 separation of 4.5 cm and a capacitor of 1 F in series with the voltage source.  
The simulation
grid is uniform and consists of 1000 cells. The electron time
step is  $3.68 \times 10^{-11}$ s.  The simulation was run
for $5.5 \times 10^6$ time steps or 2750 rf cycles. 
For the heavy particles we use  sub-cycling and the heavy particles 
are advanced every  16 electron time steps \citep{kawamura00:413}.  We assume that the 
initial density profiles are parabolic.

\subsection{The oxygen discharge}

We consider an oxygen discharge that consists of:  electrons, 
 the ground state oxygen molecule O$_2($X$^3 \Sigma_{\rm g}^-)$,
the metastable oxygen  molecules O$_2$(a$^1\Delta_{\rm g}$),
and O$_2$(b$^1\Sigma_{\rm g}^+$),
the ground state  oxygen atom O($^3$P), the metastable oxygen atom O($^1$D), 
the negative oxygen ion O$^-$, and  the positive oxygen  ions O$^+$ and O$_2^+$.  
The basic reaction set and cross sections included in the {\tt oopd1} code for the oxygen
discharge excluding the metastable states is discussed in an earlier work and includes reactions among 
electrons, ground state molecules, ground state atoms, negative ions O$^-$, and positive ions O$^+$ and O$^+_2$
 \citep{gudmundsson13:035011}. 
Similarly, the  reactions and cross sections 
involving the metastable oxygen atom O($^1$D) and the metastable oxygen molecule
O$_2$(a$^1\Delta_{\rm g}$) are discussed by Gudmundsson and Lieberman \citep{gudmundsson15:035016}. The reactions 
involving the metastable singlet molecule O$_2$(b$^1\Sigma_{\rm g}^+$), that are included in the model,
and the energy dependent secondary electron emission yields are
 discussed by Hannesdottir and Gudmundsson \citep{hannesdottir16:055002}.  
The combined reaction set discussed in these three publications
 \citep{gudmundsson13:035011,gudmundsson15:035016,hannesdottir16:055002}
is referred to as the full reaction set in the 
discussion below.
 The reactions explored here in particular are detachment by the singlet
metastable molecule O$_2$(a$^1\Delta_{\rm g}$)
\[
\mathrm{O}^- +  \mathrm{O}_2(\mathrm{a}^1\Delta_{\rm g}) \longrightarrow
 \mathrm{products}
\]
and detachment by the metastable  O$_2$(b$^1\Sigma_{\rm g}^+$) 
\[
\mathrm{O}^- +  \mathrm{O}_2(\mathrm{b}^1\Sigma_{\rm g}^+) \longrightarrow
 \mathrm{O}_2(\mathrm{X}^3 \Sigma_{\rm g}^-) +  \mathrm{O}(^3\mathrm{P}) + {\rm e}.
\]
For the detachment by
 the metastable  O$_2$($\mathrm{a}^1\Delta_{\rm g}$)  we use the rate coefficient
measured at 400 K of $1.5 \times 10^{-16}$ m$^{2}$/s by Midey et al.~\citep{midey08:3040} to estimate the cross
section, which is allowed to fall as $\sim 1/\sqrt{E}$, where $E$ is the energy, to 184 meV and then
take a fixed value of $5.75 \times 10^{-20}$ m$^2$.
Also we assume that  detachment by
 the metastable  O$_2$($\mathrm{a}^1\Delta_{\rm g}$) leads to the formation of O($^3$P) +
 O$_2(\mathrm{X}^3 \Sigma_{\rm g}^-)$ + e, instead of O$_3$ + e and O + O$_2^-$.
The cross section for detachment from O$^-$ by O$_2$(b$^1\Sigma_{\rm g}^+$) is
estimated from the rate coefficient of $6.9 \times 10^{-16}$ m$^3$/s, 
given by \citet{aleksandrov78:806} 
and then we allow it to
fall as $\sim 1/\sqrt{E}$ for $E < 184$ meV and then remain constant.

The neutral gas density is much higher than the densities of charged species, so 
the neutral species at thermal energies (below a certain cut-off energy) are treated as a background with fixed
density and temperature and maintained uniformly in space. These neutral background species are assumed to have a
Maxwellian velocity distribution at the gas temperature (here~$\mathrm{T}_{\mathrm{n}}$ = 26 mV).
In the oxygen discharge, the
 densities of the  metastable oxygen  molecules O$_2$(a$^1\Delta_{\rm g}$) and  O$_2$(b$^1\Sigma_{\rm g}^+$), 
oxygen atom in the ground state O($^3$P) and the  metastable oxygen atom O($^1$D) 
are  much larger than the number of charged species. 
We assume that the background
consists of 4.4 \%  O$_2$(a$^1\Delta_{\rm g}$), 4.4 \% O$_2$(b$^1\Sigma_{\rm g}^+$), 90.6 \%  O$_2($X$^3 \Sigma_{\rm g}^-)$,  0.519 \% O($^3$P)
and 0.028 \% O($^1$D) as in our earlier work  \citep{hannesdottir16:055002}. 
The neutrals are treated kinetically as particles if their energy exceeds a preset threshold value.
The threshold values used here for the neutral species are listed in table \ref{simpar}.
As a neutral species hits the electrode it returns as a thermal particle with a given probability
and atoms can recombine to form a thermal molecule with the same probability.
These  wall recombination  probabilities as well as the quenching coefficients 
used for the excited neutral species are listed in table \ref{simpar}.  The choice of the 
quenching coefficients for the molecular metastables and its influence on the discharge
properties is discussed towards the end of next section.
The secondary electron emission yield for O$_2^+$-, O$^+$-ions, O$_2$ molecules and O atoms are  energy dependent and we used the fits 
for oxidized metal given elsewhere \citep{hannesdottir16:055002}.
\begin{table}[h]
\caption{The parameters of the simulation, the particle weight, the threshold above which the dynamics of the neutral particles are followed and the wall 
recombination and quenching coefficients used.}
\label{simpar}
\tabcolsep7pt\begin{tabular}{llll}
\hline
Species & particle weight & threshold  & wall quenching or   \\
        &                 &  [meV]    & recombination coefficient        \\
\hline
 O$_2(\mathrm{X}^3 \Sigma_{\rm g}^-)$     &  $5 \times 10^7$  & 500  & 1.0               \\    
 O$_2$($\mathrm{a}^1\Delta_{\rm g}$)      &  $5 \times 10^7$  & 100  &0.007  \citep{sharpless89:7947}    \\    
O$_2$($\mathrm{b}^1\Sigma_{\rm g}^+$)      &  $5 \times 10^7$  & 100  &0.1  (see text)    \\    
  O($^3$P)                          &  $5 \times 10^8$  & 500  &0.5            \\    
  O($^1$D)                          &  $5 \times 10^8$  & 50   &1.0  (0.5 recomb., 0.5 quenching)         \\ 
      O$_2^+$                       &   $10^7$          & -  &     -           \\ 
      O$^+$                         &   $10^6$          & -  &   -           \\ 
      O$^-$                         &   $5 \times 10^6$ & -  &   -           \\ 
     e                       &   $1 \times 10^8$ & -  &   -           \\ 
\hline
\end{tabular}
\end{table}


\section{RESULTS AND DISCUSSION} 

\begin{figure}[h]
\centerline{\includegraphics[width=0.90\linewidth]{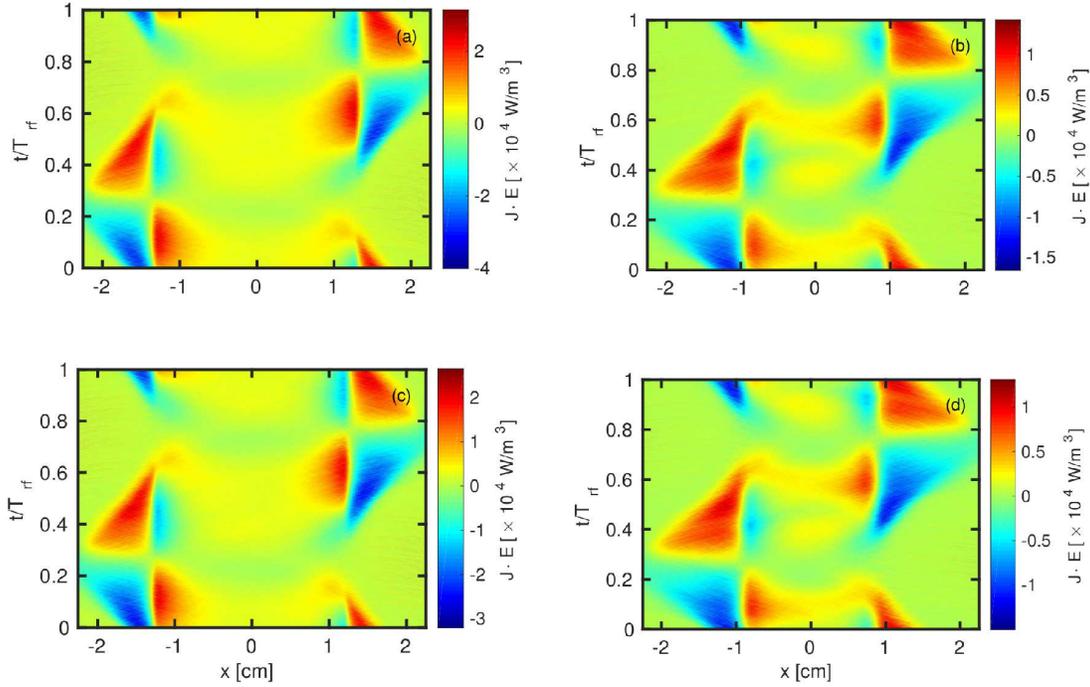}}
\vspace*{3mm}
\caption{The electron heating profile for a parallel plate capacitively coupled 
oxygen discharge at 10 mTorr with a gap separation of 4.5 cm driven by a 222 V voltage 
source at 13.56 MHz.   The four cases explored are:  (a) detachment neither by  O$_2$(a$^1\Delta_{\rm g}$)  nor  O$_2$(b$^1\Sigma_{\rm g}^+$), 
(b) only detachment by  O$_2$(b$^1\Sigma_{\rm g}^+$) included, (c)  only detachment by O$_2$(a$^1\Delta_{\rm g}$) included, 
(d) and both detachment by 
  O$_2$(a$^1\Delta_{\rm g}$)  and  O$_2$(b$^1\Sigma_{\rm g}^+$) included (full reaction set).  \label{jdotE10}} 
\end{figure}
\begin{figure}[h]
\centerline{\includegraphics[width=0.90\linewidth]{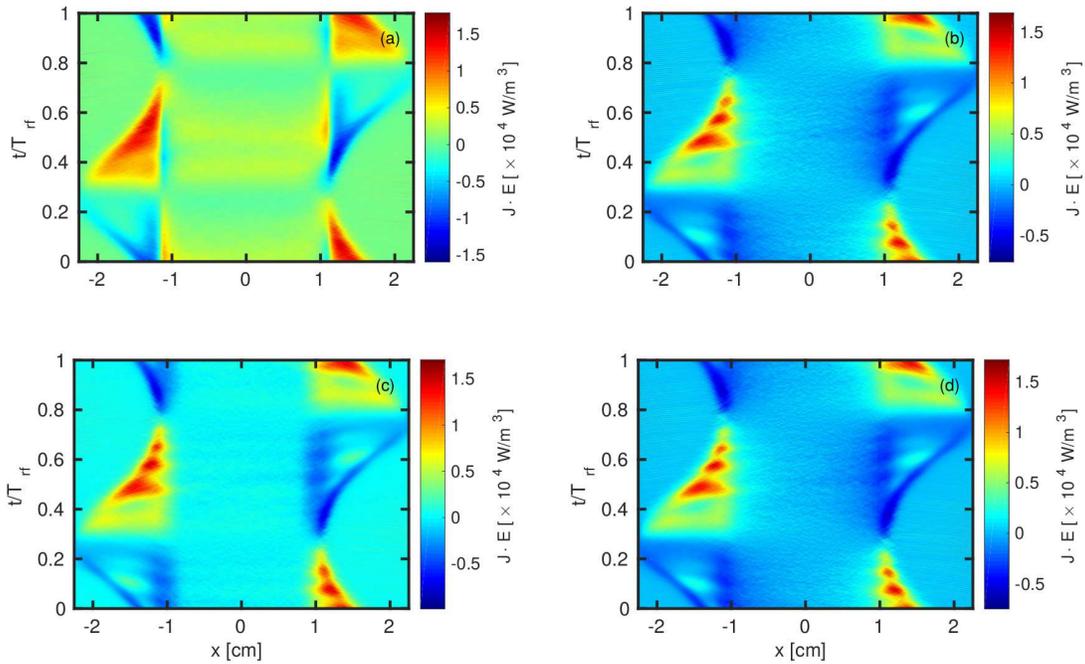}}
\vspace*{3mm}
\caption{The electron heating profile for a parallel plate capacitively coupled 
oxygen discharge at 50 mTorr with a gap separation of 4.5 cm driven by a 222 V voltage 
source at 13.56 MHz.  The four cases explored are:  (a) detachment neither by  O$_2$(a$^1\Delta_{\rm g}$)  nor  O$_2$(b$^1\Sigma_{\rm g}^+$), 
(b) only detachment by  O$_2$(b$^1\Sigma_{\rm g}^+$) included, (c)  only  detachment by O$_2$(a$^1\Delta_{\rm g}$) included, (d) 
and both detachment by 
  O$_2$(a$^1\Delta_{\rm g}$)  and  O$_2$(b$^1\Sigma_{\rm g}^+$) included (full reaction set).  \label{jdotE50}} 
\end{figure}
\begin{figure}[h]
\centerline{\includegraphics[width=0.90\linewidth]{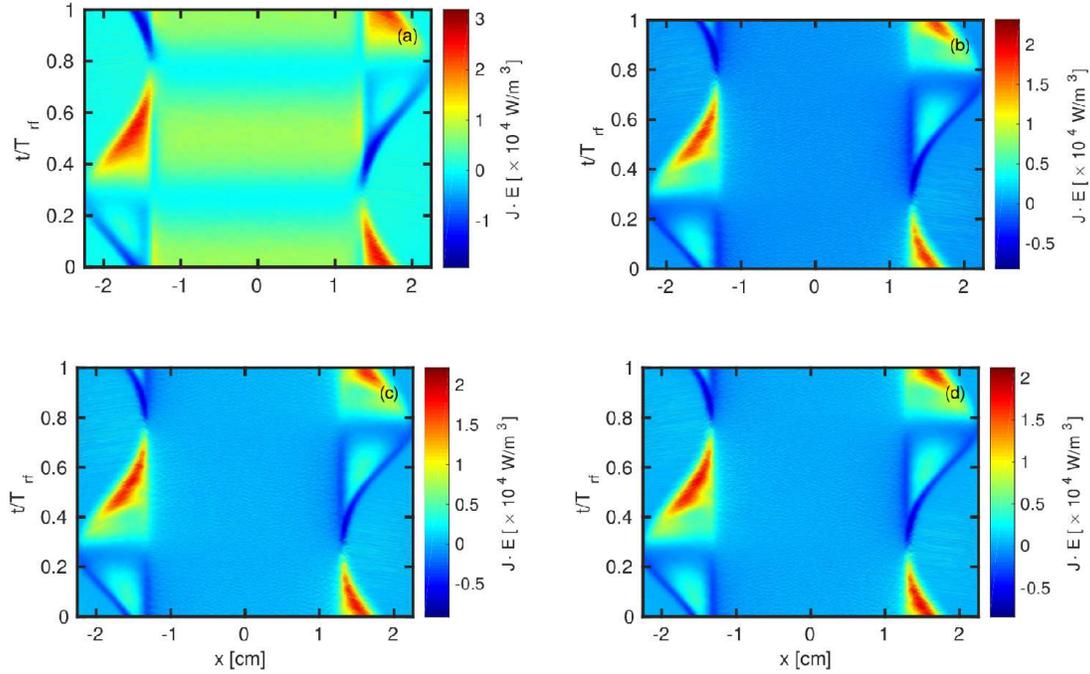}}
\vspace*{3mm}
\caption{The electron heating profile for a parallel plate capacitively coupled 
oxygen discharge at 200 mTorr with a gap separation of 4.5 cm driven by a 222 V voltage 
source at 13.56 MHz.  The four cases explored are:  (a) detachment neither by  O$_2$(a$^1\Delta_{\rm g}$)  nor  O$_2$(b$^1\Sigma_{\rm g}^+$), 
(b) only detachment by  O$_2$(b$^1\Sigma_{\rm g}^+$) included, (c)  only detachment by O$_2$(a$^1\Delta_{\rm g}$) included, 
(d) and both detachment by 
  O$_2$(a$^1\Delta_{\rm g}$)  and  O$_2$(b$^1\Sigma_{\rm g}^+$) included (full reaction set).  \label{jdotE200}} 
\end{figure}

The electron power absorption  is calculated as 
 ${\bf J}_{\rm e} \cdot {\bf E}$, where ${\bf J}_{\rm e}$
and ${\bf E}$ are the spatially and temporally varying electron current
density and electric field, respectively.  
The spatio-temporal behavior of the electron power absorption,  for pressures 10, 50 and 200 mTorr, is shown in figures \ref{jdotE10} -- \ref{jdotE200}, respectively,
 for four cases, (a) when the detachment by 
both the metastable states  O$_2$(a$^1\Delta_{\rm g}$) and  O$_2$(b$^1\Sigma_{\rm g}^+$) are neglected, (b) when  detachment by O$_2$(a$^1\Delta_{\rm g}$) is neglected and  by
O$_2$(b$^1\Sigma_{\rm g}^+$) is included, (c) when  detachment by O$_2$(a$^1\Delta_{\rm g}$) is included and  by O$_2$(b$^1\Sigma_{\rm g}^+$) is neglected,  and  (d) when the detachments by 
both the metastable states  O$_2$(a$^1\Delta_{\rm g}$) and  O$_2$(b$^1\Sigma_{\rm g}^+$) are included in the simulations (the full reaction set). For each of the figures the abscissa covers 
the whole inter-electrode gap, from the powered electrode on the left hand side to the grounded electrode on the right hand side.  Similarly the ordinate covers the full rf cycle.   
The four figures grouped together for each pressure value can have  different magnitude scales. Therefore, 
there can be differences in the four figures, not only qualitatively but also quantitatively. When comparing the figures for each pressure
 be aware of such subtle differences.
Figure  \ref{jdotE10} (a) shows the spatio-temporal behavior of the electron power absorption at 10 mTorr  when the detachments by 
both the metastable states  O$_2$(a$^1\Delta_{\rm g}$) and  O$_2$(b$^1\Sigma_{\rm g}^+$) are neglected in the simulation.  The most significant heating is in the sheath region,
however there is a significant contribution to the heating in the bulk region.  Both energy gain (red and yellow areas) and energy loss (dark blue areas) are evident.
 As detachment by  O$_2$(b$^1\Sigma_{\rm g}^+$) only  is added to the simulation in figure 
 \ref{jdotE10} (b) the  spatio-temporal heating   structure is maintained but the sheath width increases.  Note that there is also a change in the scale when 
 detachment by  O$_2$(b$^1\Sigma_{\rm g}^+$)  is added, the peak electron heating in the sheath region decreases.  Adding detachment by  O$_2$(a$^1\Delta_{\rm g}$) only does not
lead to decreased sheath with as seen in figure  \ref{jdotE10} (c).   Including both the detachment processes    (the full reaction set)  shows 
 very similar  spatio-temporal heating   structure as when only   O$_2$(b$^1\Sigma_{\rm g}^+$) is included in the simulation.   In all cases the electron heating is most significant during the 
the sheath expansion phase at each electrode (the red areas).  
We also observe electron heating during the sheath collapse on the bulk side of the sheath edge while there is cooling (electrons loose energy) on the
electrode side (the lower left hand corner and upper center on the right hand side).  
This is in accordance to what is commonly observed in single frequency capacitively coupled discharges
\citep{wood91t}.  Setting the electron emission coefficient 
to zero gives almost exactly the same spatio-temporal heating structure for all the four cases (not shown).
The spatio-temporal heating   structure at 50 mTorr is shown in figure \ref{jdotE50}.  When detachment by both the metastable states is neglected
there is significant heating in the electronegative core  as seen in figure \ref{jdotE50} (a) (green and yellow areas).  When detachment by the singlet state O$_2$(a$^1\Delta_{\rm g}$) only is included,
the heating in the electronegative core decreases as seen in figure   \ref{jdotE50} (c)   but when detachment by the singlet state   O$_2$(b$^1\Sigma_{\rm g}^+$) only 
is included the heating in the electronegative core decreases more or to almost zero as seen in figure   \ref{jdotE50} (b).  Including both the detachment processes  (the full reaction set)  shows 
almost zero heating in the electronegative core as seen in figure   \ref{jdotE50} (d).  In all cases the electron heating is most significant during the 
the sheath expansion phase at each electrode (the red areas).  We note oscillations in the heating like those reported by Vender and Boswell \citep{vender92:1331} as we add
detachment by the singlet metastable molecules to the simulations.  We also note that cooling in the sheath region during the sheath collapse is always apparent 
(note the different scale).
At 200 mTorr when detachment by both the metastable states is neglected
there is still a significant heating in the electronegative core as seen in figure \ref{jdotE200} (a).
When detachment by either   O$_2$(a$^1\Delta_{\rm g}$) or    O$_2$(b$^1\Sigma_{\rm g}^+$) is included in the simulation
the heating in the electronegative core drops to zero, as seen in figures  \ref{jdotE200} (c) and  \ref{jdotE200} (b), respectively.  
As a consequence including both the detachment processes shows  almost zero heating in the electronegative core as seen in figure   \ref{jdotE200} (d). 
 In all cases the electron heating is most significant during the 
the sheath expansion phase at each electrode (the red areas).
  Greb et al.~\citep{greb15:044003} 
find that at 300 mTorr ohmic (collisional) heating is the dominant heating mechanism far exceeding collisionless heating.  
They see that collisional electron heating has the highest contribution close to the sheath edge, during the sheath expansion and sheath collapse phases.  They also find that collisionless electron heating has higher contribution at the sheath edge during 
sheath expansion  than collisional heating.  For collisionless heating they find high negative electron heating (or cooling) at the sheath edge during the sheath collapse phase.

\begin{figure}
\includegraphics[width=0.470\linewidth]{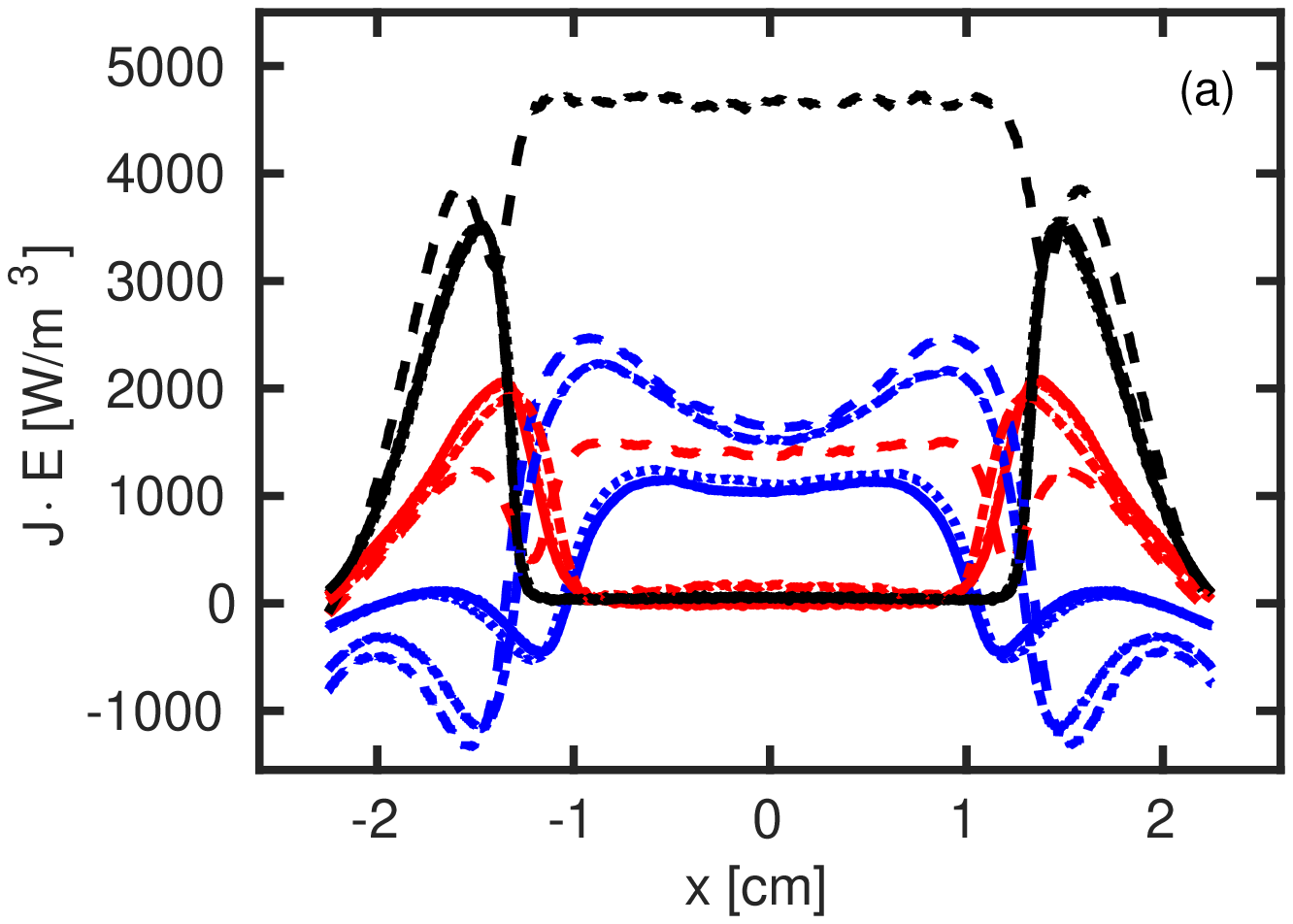} \hspace{0.6 cm}

\includegraphics[width=0.430\linewidth]{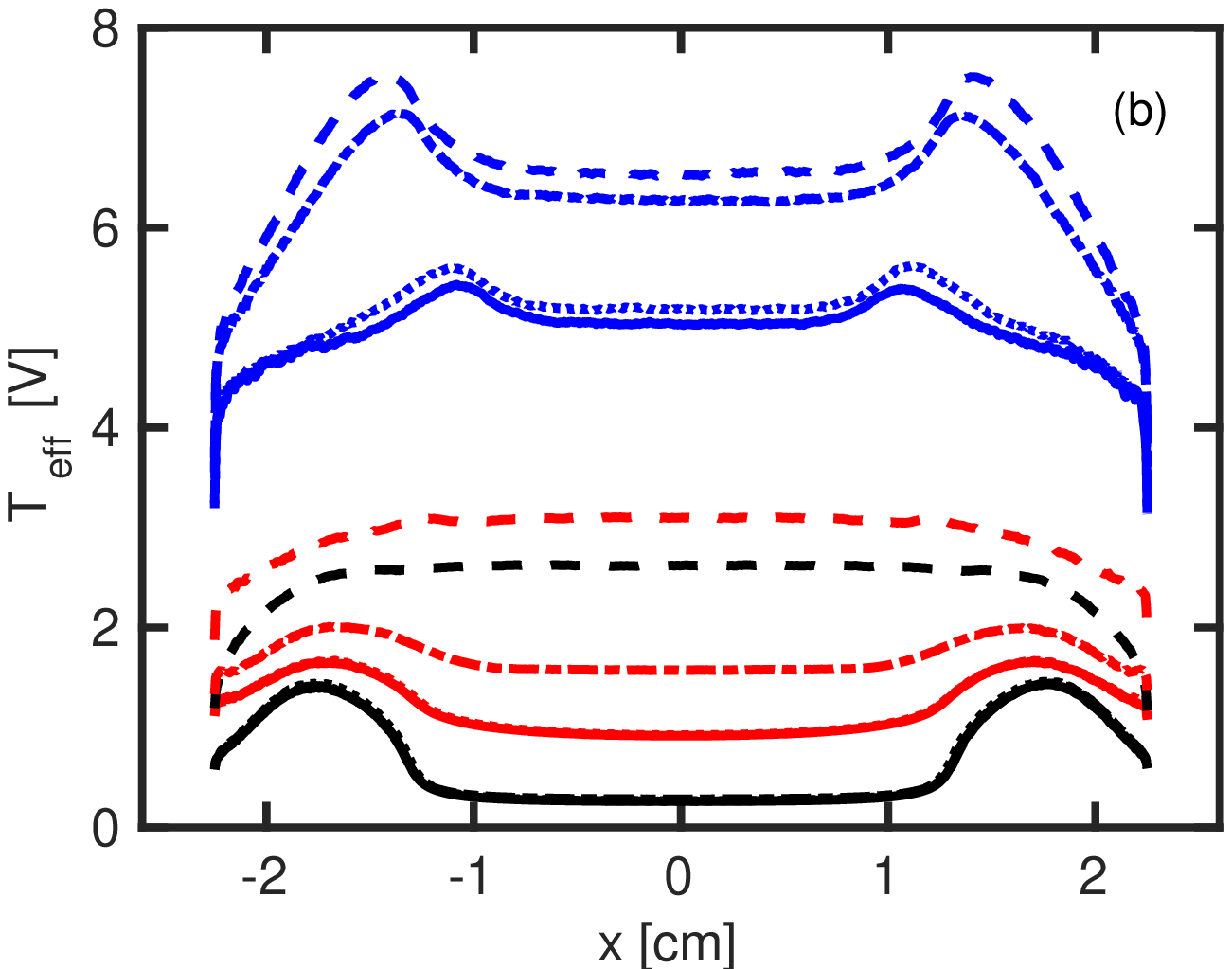}
\vspace*{3mm}
\caption{(a) The electron heating rate profile and (b)  the effective electron temperature profile for a parallel plate capacitively coupled 
oxygen discharge at 10 ({\color{blue}blue}), 50 ({\color{red}red}) and 200 (black) mTorr with a gap separation of 4.5 cm driven by a 222 V voltage  source at 13.56 MHz. 
 The four cases explored are:  detachment neither by  O$_2$(a$^1\Delta_{\rm g}$)  nor  O$_2$(b$^1\Sigma_{\rm g}^+$) (dashed line), 
only detachment by  O$_2$(b$^1\Sigma_{\rm g}^+$)  included (dotted line),  only detachment by  O$_2$(a$^1\Delta_{\rm g}$) included (dash-dot line), and both detachment by 
  O$_2$(a$^1\Delta_{\rm g}$)  and  O$_2$(b$^1\Sigma_{\rm g}^+$) included (full reaction set) (solid line).
\label{JdotE_all}} 
\end{figure}

Figure \ref{JdotE_all} (a) shows the time averaged  electron heating profile at 10, 50 and 200 mTorr for the four cases explored here.  At 10 mTorr the electron
heating is mainly ohmic heating in the electronegative core.  This bulk heating decreases only slightly when detachment is  by O$_2$(a$^1\Delta_{\rm g}$) only is
added to the reaction set, but decreases more significantly  when
 detachment by O$_2$(b$^1\Sigma_{\rm g}^+$) only is included in the reaction set.
Having the full reaction set in the simulation gives almost the same electron heating rate
profile as when   when
 detachment by O$_2$(b$^1\Sigma_{\rm g}^+$) only is included.  Note that when we neglect the detachment by the metastables
there is effective cooling in the sheath region during one period as seen in figure \ref{JdotE_all} (a). This is due to very strong cooling in the sheath collapse phase as seen in figure \ref{jdotE10} (a). The strong heating in the sheath region during the sheath expansion of the rf cycle is not significant enough to overcome this cooling during one period.  When both detachment processes are included these cooling and heating phases balance and there is almost no
effective heating in the sheath regions.     At 50 mTorr the ohmic heating in the
electronegative core is significant when detachment by the singlet metastables is neglected.  However, ohmic heating in 
the electronegative core drops to almost zero when either   detachment by O$_2$(a$^1\Delta_{\rm g}$) or  O$_2$(b$^1\Sigma_{\rm g}^+$) is included in the simulation.  
Detachment by  O$_2$(b$^1\Sigma_{\rm g}^+$) only has slightly stronger influence on the ohmic heating in the electronegative core.  At 200 mTorr  there is a significant  ohmic heating in the
electronegative core when detachment by both the singlet metastables is neglected and it drops to almost zero when either detachment by O$_2$(a$^1\Delta_{\rm g}$) or  
O$_2$(b$^1\Sigma_{\rm g}^+$) is included in the simulation.  The peak electron heating in the sheath region is higher at 200 mTorr than at 50 mTorr.
The same is seen when exploring the effective electron temperature profile shown in figure  \ref{JdotE_all} (b).
  The effective electron temperature profile 
changes significantly when detachment by singlet metastables is added to the reaction set.  At 10 mTorr detachment by O$_2$(a$^1\Delta_{\rm g}$) only 
has small effect on the effective electron temperature while  detachment by  O$_2$(b$^1\Sigma_{\rm g}^+$) only decreases the effective electron temperature significantly.  At 50 mTorr 
detachment by both  O$_2$(a$^1\Delta_{\rm g}$)  and  O$_2$(b$^1\Sigma_{\rm g}^+$) lowers the  effective electron temperature but detachment by   O$_2$(b$^1\Sigma_{\rm g}^+$)  has more influence.  
At 200 mTorr including either one of the  detachment processes is enough to  get the same value of the effective electron temperature as when the full reaction set is used in the simulations. 
This low effective electron
temperature in the discharge center indicates that the heating of the electrons is effective in
the sheath region, but the ohmic (or collisional) heating in the discharge center is ineffective. The effective 
electron temperature when the full reaction set is assumed in the simulations is
 in accordance with the measurements of Kechkar \citep{kechkar15t} which find $T_{\rm eff} < 1$ eV in
the pressure range 50 - 200 mTorr in a slightly asymmetric discharge in pure oxygen.  Furthermore,
 Pulpytel et al.~\citep{pulpytel07:073308} measured the effective electron temperature in an Ar/O$_2$ (1:3)
mixture at 110 mTorr and voltages above 300 V to be around 1 eV.

\begin{figure}
\includegraphics[width=0.43\linewidth]{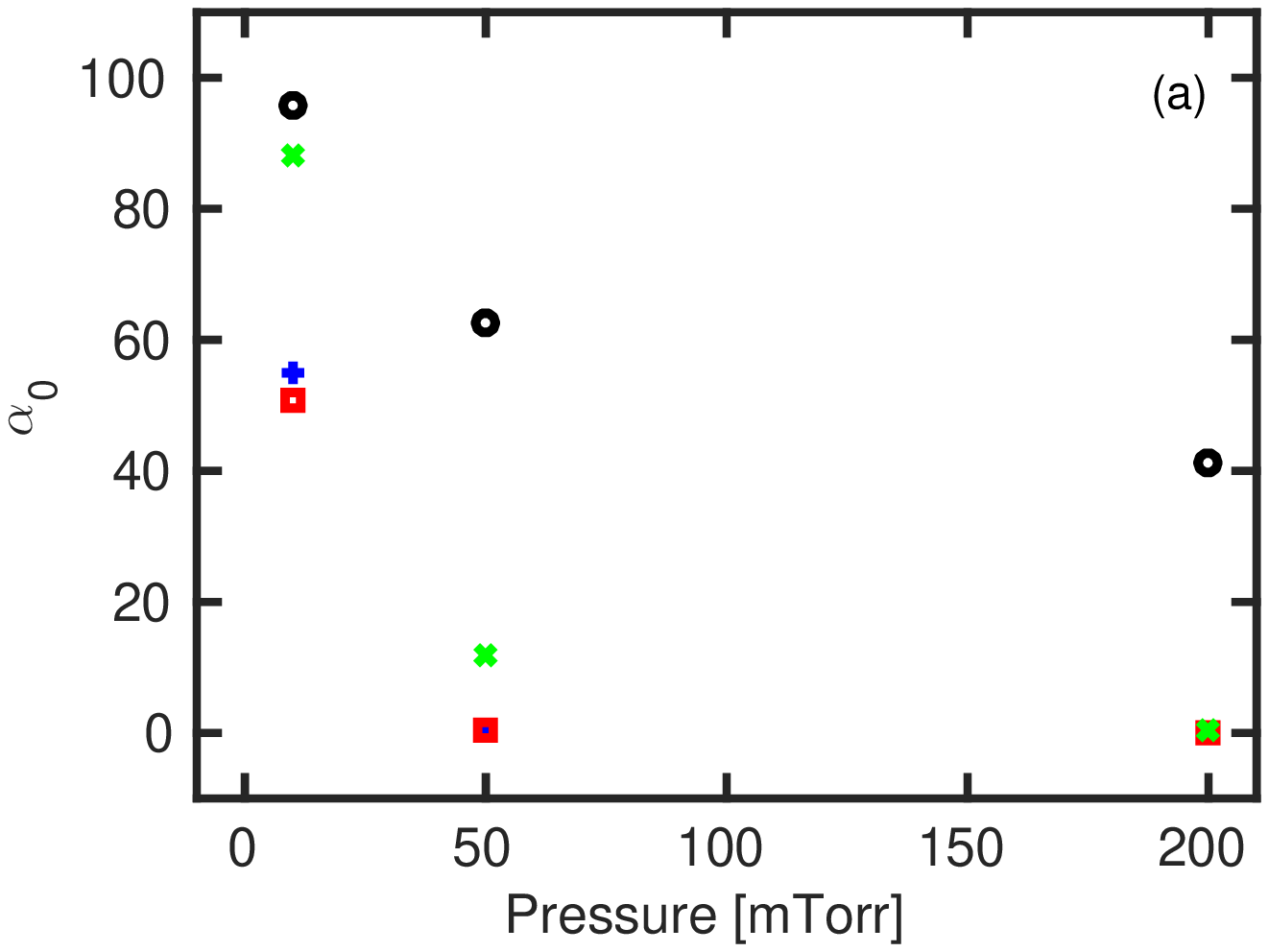}  \hspace{0.6 cm}

\includegraphics[width=0.43\linewidth]{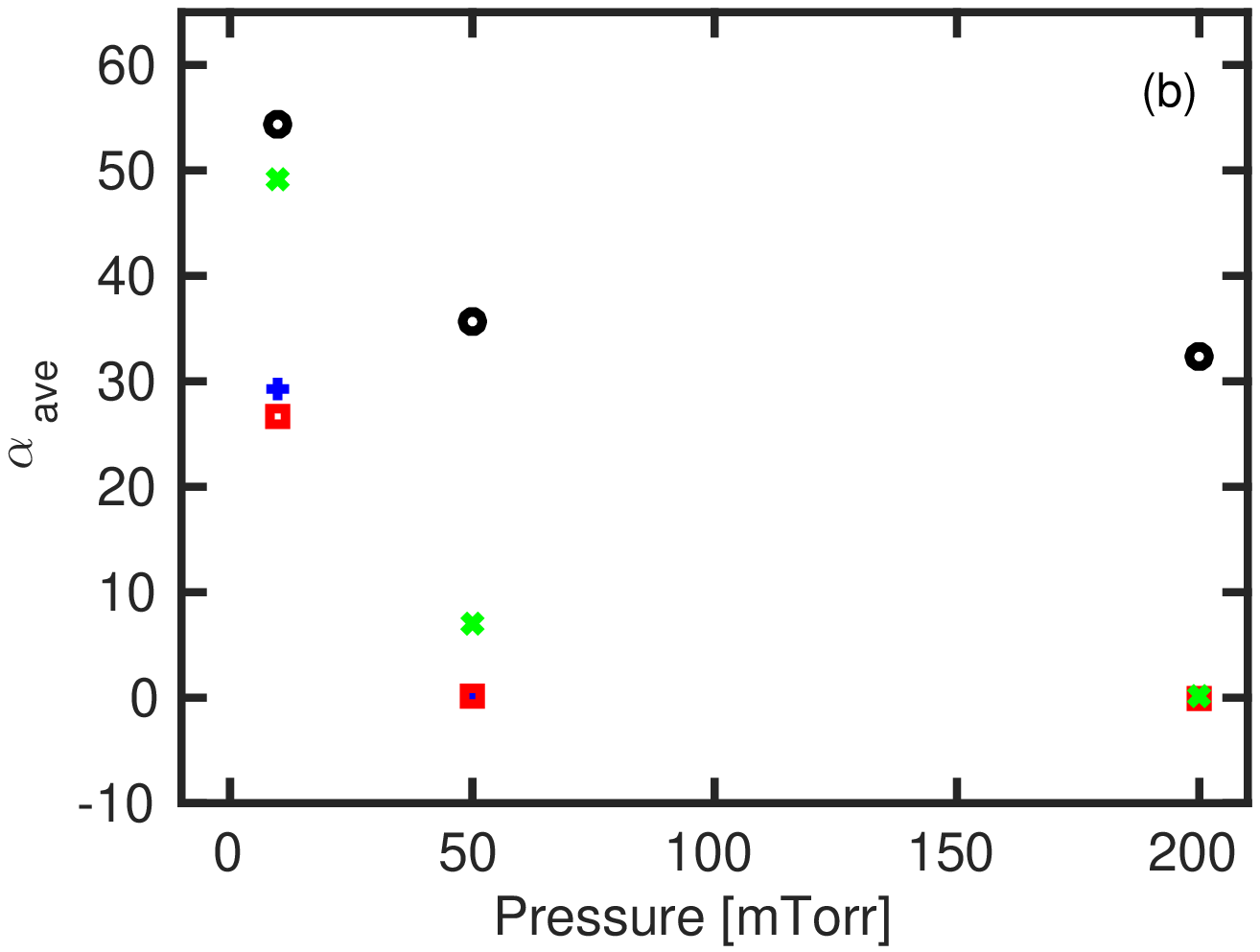}
\vspace*{3mm}
\caption{ (a) The center electronegativity and (b) the average electronegativity as a function of pressure for a parallel plate capacitively coupled 
oxygen discharge at 50 mTorr with a gap separation of 4.5 cm driven by a 222 V voltage 
source at 13.56 MHz.  The four cases explored are:  detachment neither by  O$_2$(a$^1\Delta_{\rm g}$)  nor  O$_2$(b$^1\Sigma_{\rm g}^+$)   ({\color{black} {\bf o}}), 
only detachment by  O$_2$(a$^1\Delta_{\rm g}$) included ({\color{green} ${\bf \times}$}),  only detachment by O$_2$(b$^1\Sigma_{\rm g}^+$) included ({\color{blue} {\bf +}}), and both detachment by 
  O$_2$(a$^1\Delta_{\rm g}$)  and  O$_2$(b$^1\Sigma_{\rm g}^+$) included (full reaction set) ({\color{red}${\bf \Box}$}).
 \label{alpha0}},  
\end{figure}
Figure \ref{alpha0} (a) shows the center electronegativity $\alpha_0 = n_{-0}/n_{\rm e0}$ where $ n_{-0}$ is the center negative ion density and $n_{\rm e0}$
is the center electron density.  The negative ions are confined to the plasma bulk due to a potential well, and the negative ion density has a peak near the 
discharge center. We see that the center electronegativity decreases with increasing discharge pressure.  
Stoffels et al.~\citep{stoffels95:2425} measure the electronegativity to be in the range 5 -- 10 and find it to 
decrease with increased pressure.  This is somewhat higher electronegativity than what we show here.
The pressure dependence contradicts the findings reported by Bera et al.~\citep{bera11:1027}
which apply a PIC/MCC hybrid model to a low pressure  oxygen discharge, and observe an increase in electronegativity with increased pressure. As seen in figure  \ref{alpha0} (a) the electronegativity decreases as we add detachment 
processes to the discharge model. Adding detachment by  O$_2$(b$^1\Sigma_{\rm g}^+$) has more significant influence on the electronegativity than adding detachment by
 O$_2$(a$^1\Delta_{\rm g}$). The average electronegativity is given as
\begin{equation}
\alpha_{\rm ave} = \frac{\int_0^d n_-(x) dx}{\int_0^d n_{\rm e}(x) dx}
\end{equation}
 where $n_-$ is the O$^-$-ion density, $n_{\rm e}$ is the electron density and $d$ is the electrode separation.   
We see that the average electronegativity shown in figure \ref{alpha0} (b)   
is always somewhat lower than the center electronegativity.    
Earlier we explored the influence of increasing the singlet molecular metastable O$_2$(a$^1\Delta_{\rm g}$) density on the 
electron heating rate.  We
find that increasing the partial pressure of the metastable molecule O$_2$(a$^1\Delta_{\rm g}$)
in the background does not have much influence on the electron heating except that 
the ohmic heating in the bulk is closer to zero and the electronegativity is slightly lower
when 8.8 \% of the discharge pressure is due to the metastable molecule 
O$_2$(a$^1\Delta_{\rm g}$) \citep{gudmundsson15:153302}.
Similarly  increasing the partial pressure of O$_2$(b$^1\Sigma_{\rm g}^+$) decreases the electronegativity even further 
\citep{hannesdottir16:055002}.  But for a  change in partial pressure of O$_2$(b$^1\Sigma_{\rm g}^+$)  from  2.2 \% to 4.4 \% 
there  is not a  significant change in electronegativity.  
It has been demonstrated by using a 1D fluid model that the electronegativity depends strongly 
on the  O$_2$(a$^1\Delta_{\rm g}$)     surface quenching probability \citep{greb15:044003}.  
For this current study we use a quenching probability of 0.007 estimated by Sharpless and Slanger \citep{sharpless89:7947} for iron.  Their estimate for aluminum is $< 10^{-3}$.   
Greb et al.~\citep{greb15:044003} argue that increased quenching coefficient leads to 
decreased  O$_2$(a$^1\Delta_{\rm g}$) density and thus decreased detachment by the   O$_2$(a$^1\Delta_{\rm g}$) state and thus higher negative ion density.  For iron Ryskin and Shub \citep{ryskin81:41} report a value of 0.0044 and, for aluminum,  $5 \times 10^{-5}$.  Aluminum electrodes would therefore lead to higher singlet metastable densities and lower electronegativity. 
We have seen in global model studies that wall quenching can be the main loss mechanism for the singlet metastable
state   O$_2$(b$^1\Sigma_{\rm g}^+$) \citep{toneli15:325202}.  In these studies we assumed the quenching coefficient to be 0.1, which is the same value as that assumed in this current study.  This assumption is based on the suggestion that the quenching coefficient  for the 
b$^1\Sigma_{\rm g}^+$ state is about two orders of magnitude larger than that for the a$^1\Delta_{\rm g}$ state \citep{obrien70:3832}.

\section{CONCLUSION}  

We used the one-dimensional object-oriented particle-in-cell Monte Carlo collision code {\tt oopd1}
to explore the  spatio-temporal evolution of the electron heating mechanism with pressure 
in  a capacitively coupled oxygen discharge including and excluding the detachment by the singlet molecular metastable states.  
We find that  detachment by the singlet metastable state O$_2$(b$^1\Sigma_{\rm g}^+$) has significant effect while
  O$_2$(a$^1\Delta_{\rm g}$) has somewhat less significant influence on the discharge properties.  We conclude that
 it is essential to include detachment by both the metastable states. 
Including the detachment processes has a strong influence on the effective electron temperature and electronegativity in the oxygen discharge, 
both of these parameters are significantly lower when the detachment processes are 
included in the simulation.   However, it has to be kept in mind that the wall quenching probability for each of the molecular metastable state has a 
significant influence on the overall discharge and the actual values of the electronegativity and the effective electron temperature.  
  At 10 mTorr, using the full reaction set, the time averaged 
 electron heating displays mainly contribution from 
 ohmic heating in the plasma bulk (the electronegative core) and at higher pressures there is no ohmic heating in the plasma bulk and electron heating in the sheath regions dominates.


\section{ACKNOWLEDGMENTS}
This work was partially supported by the
 Icelandic Research Fund Grant
Nos.~130029 and  163086, and 
 the Swedish Government Agency for Innovation Systems (VINNOVA) contract no. 2014-04876.


%

\end{document}